\documentclass[aps,pra,reprint,superscriptaddress,amsmath,showpacs,amssymb,floatfix]{revtex4-1}
\usepackage{bm}
\usepackage[dvips,final]{graphicx}
\usepackage{amsmath,amssymb}

\date{\today} 

\newcommand{\be}{\begin{equation}}
\newcommand{\ee}{\end{equation}}
\newcommand{\BM}{\begin{pmatrix}}
\newcommand{\EM}{\end{pmatrix}}

\bmdefine{\bx}{x}
\bmdefine{\by}{y}
\bmdefine{\bz}{z}
\bmdefine{\bu}{u}
\newcommand{\kone}{\kappa _1}
\newcommand{\ktwo}{\kappa _2}

\newcommand{\pt}{\frac{\partial }{\partial t}}
\newcommand{\jbar}{\overline{j}}

\newcommand{\psij}{\psi _j(\bx,t)}
\newcommand{\psijbar}{\psi _{\jbar}(\bx,t)}

\newcommand{\xij}{\xi _j(\bx)}

\newcommand{\kj}{\kappa _j}

\begin{document}
\title{Analytical study on parameter regions of dynamical
instability for two-component Bose--Einstein condensates with coaxial quantized vortices}

\author{M.~Hoashi}
\altaffiliation[Present address: ]
{Fujitsu limited, 
1-17-25 Shin-kamata, Ohta-ku, Tokyo 144-8588, Japan\\
hoashi.masaki@jp.fujitsu.com}
\affiliation{Department of Electronic and Physical Systems, Waseda
University, Tokyo 169-8555, Japan} 
\author{Y.~Nakamura}
\email{yusuke.n@asagi.waseda.jp}
\affiliation{Department of Electronic and Physical Systems, Waseda
University, Tokyo 169-8555, Japan} 

\author{Y.~Yamanaka}
\email{yamanaka@waseda.jp}
\affiliation{Department of Electronic and Physical Systems, Waseda
University, Tokyo 169-8555, Japan} 

\begin{abstract} 
The dynamical instability of weakly interacting 
two-component Bose--Einstein condensates with coaxial quantized vortices is analytically investigated
in a two-dimensional isotopic harmonic potential. We examine whether 
complex eigenvalues appear on the Bogoliubov--de Gennes equation, implying dynamical instability. 
Rather than solving the Bogoliubov--de Gennes equation numerically, we rely on a perturbative expansion with respect to the coupling constant which enables a simple, analytic approach.
For each pair of winding numbers and for each magnetic quantum number, the ranges of 
inter-component coupling constant where the system is dynamically unstable
are exhaustively obtained. Co-rotating and counter-rotating systems show distinctive
behaviors. The latter is much more complicated than the former with respect to dynamical 
instability, particularly because radial excitations contribute to complex eigenvalues in counter-rotating systems.
\end{abstract}

\pacs{03.75.Kk, 03.75.Lm, 67.85.Fg}

\maketitle

\section{INTRODUCTION}
Dynamical instability is one of the most interesting phenomena in Bose--Einstein condensates of cold atomic gases. This instability is observed experimentally in various situations whose typical examples include the splitting of a multiply quantized vortex \cite{Shin} and the decaying of a condensate flowing in an optical lattice \cite{Fallani}. Theoretically, the dynamics of condensates are well described by the time-dependent Gross--Pitaevskii (TDGP) equation \cite{Dalfovo}, and  theoretical studies solving the TDGP equation successfully explained the experiment of vortex splitting \cite{Huhtamaki,Munoz}. When judging whether the condensate is dynamically unstable, we may employ the Bogoliubov--de Gennes (BdG) equation \cite{Bogoliubov, deGennes, Fetter}, which is obtained by linearizing the TDGP equation. The BdG equation is a non-hermetian eigenvalue problem, giving complex eigenvalues as well as real eigenvalues, and we interpret the presence of complex eigenvalues as an indication of dynamical instability. By solving the BdG equation under given physical conditions, we can find regions of parameters in which the system is dynamically unstable. 

The dynamical instability of a multiply quantized vortex in a single component system has been widely investigated. In this study, we consider multi-component systems with quantized vortices  because understanding the instabilities in these systems is a difficult problem.  Previous works on this matter numerically solve the differential equations (see Refs.~\cite{Skryabin, Ishino, Wen}). The inter-component interaction and mutual influence between multiple vortices make the dynamical behavior of this system diverse and nontrivial.  Therefore, it is not practical to solve the BdG equation numerically in the entire parameter space or to find all parameter regions of the dynamical instability. Such an exhaustive numerical study of the BdG equation on dynamically unstable regions is not easy even for a single component system with a multiply quantized vortex  because some regions may be too small to be identified \cite{Kawaguchi}. 

The general properties of the BdG equation are well investigated \cite{Skryabin, Kawaguchi, Mine}. To address its non-hermiticity, an inner product must be introduced with an indefinite metric, which guarantees orthonormality to the eigenfunction set. Then, eigenfunctions belonging to real eigenvalues are classified according to the sign of its squared norm into positive- and negative-norm eigenfunctions. On the other hand, the squared norm of an eigenfunction with a complex (non-real) eigenvalue is always zero. The degeneracy between positive- and negative-norm eigenfunctions, a kind of resonance, has been shown both numerically and analytically to be necessary for the emergence of complex eigenvalues \cite{Lundh, Taylor, Skryabin, Kawaguchi}. In our previous study \cite{Nakamura1}, we proposed a systematic method based on perturbation theory to find parameter regions in which complex eigenvalues emerge, namely regions where the system is dynamically unstable, starting from regions without complex eigenvalues. Because this method is simple in essence, it can be extended to multi-component systems.

The aim of this paper is to investigate the dynamical instability of two-component systems with two quantized vortices whose cores overlap according to the method in Ref.~\cite{Nakamura1}. Determining whether such systems are unstable by solving the TDGP and BdG equations demands a heavy load of numerical calculations. In our method, we analytically solve algebraic equations, which are much simpler than the differential equations, and we can cover a wide area of parameters to exhaustively determine regions of dynamical instability without overlooking small regions. The only restrictions of our current study are that the coupling constants of the respective self-interactions and the inter-component interaction are assumed to be so small that the perturbative approach with respect to these coupling constants is allowed.

In Sect.~\ref{sec-GPBdG}, a general formulation of the TDGP and BdG equations is reviewed for the two-dimensional, two-component condensate system trapped by a harmonic potential. We also review our analytic method based on the perturbation method in Ref.~\cite{Nakamura1}, originally for a single component system, and consider its extension to a multi-component system in Sect.~\ref{sec-General}. There, we emphasize that the degeneracy between the unperturbed positive- and negative-norm eigenstates is a necessary prerequisite to the emergence of complex eigenvalues. Section~\ref{sec-Application} is the main part of this paper in which the formulations in the preceding sections are applied to a trapped two-component system with two coaxial quantized vortices. The unperturbed states are those with the vanishing intra- and inter-component coupling constants. A summary is given in Sect.~\ref{sec-Summary}.

\section{GENERAL FORMULATION OF GROSS--PITAEVSKII AND BOGOLIUBOV--DE GENNES EQUATIONS FOR TWO COMPONENT CONDENSATE SYSTEM}\label{sec-GPBdG}

We consider two-component condensates in the $x$-$y$ plane at zero temperature, 
trapped by a two-dimensional isotropic
harmonic potential with trap frequency $\omega$. This condensate system can be realized
as the limit of pancake-shaped condensates in a three-dimensional cylindrical harmonic
potential with a very large trap frequency along the $z$-axis. 
The system is characterized by the order parameters, denoted 
by $\psi_j (\bx,t)$
($j=1,2$ and ${\bm x}=(x,y)$), which satisfy the coupled TDGP equations,
\begin{align}
&i{\frac{\partial }{\partial t}} \psi_j (\bx,t) 
=\Bigl\{ h_0 -\mu _j \nonumber \\
& \quad +gN\beta _{jj}|\psi_j (\bx,t)| ^2
+gN\beta _{j\jbar }
\left|\psi_{\jbar} (\bx,t) \right |^2 \Bigr\} \psi_j (\bx,t) \, .\label{TDGP} 
\end{align}
Here, we use the notation of $h_0=-\nabla ^2/2m + m\omega^2 (x^2+y^2)/2$ and $\jbar=2,1$ for $j=1,2$\,, and 
$\mu_j$ stands for the chemical potential of each component $j$\,.
Throughout this paper, $\hbar$ is set to unity.
For simplicity, the masses and the condensate populations of the two species are taken to be the same and are
denoted by $m$ and $N$, respectively.
All interactions are assumed to be represented by two-body contact-type
potentials, and the three independent coupling constants are $g\beta_{11}$\,, 
$g\beta_{22}$\,, and $g\beta_{12}=g\beta_{21}$\,. 
The order parameters are normalized as 
\begin{align}
\int\!dxdy \, \left| \psi_j (\bx,t) \right|^2= 1 \,.
\end{align}
For the stationary Gross--Pitaevskii equations, the solutions 
are represented by $\xi _j(\bx)$, 
\begin{align}
\left(h_0-\mu _j+gN\beta _{jj} |\xi _j(\bx)|^2+
gN\beta _{12}|\xi _{\jbar}(\bx) |^2\right)\xi _j(\bx)=0 \,.
\label{stationaryGP}
\end{align}

We suppose time evolution of the order parameters that slightly deviate
from $\xi(\bx)$, {\it i.e.}, 
$\psi_j (\bx,t) =\xi _j(\bx)+ \delta \psi_j (\bx,t)$\,. Substituting these parameters
into Eq.~(\ref{TDGP}) and linearizing the TDGP equations with respect to
$\delta  \psi_j$, we obtain
\begin{align}
&i\pt \delta\psij =\Bigl\{ h_0-\mu _j
\nonumber \\
&\qquad +2gN\beta _{jj}|\xi_j| ^2+gN\beta _{12}|\xi_{\jbar}|^2
 \Bigr \}\delta\psij \nonumber \\
&\qquad+gN\beta _{jj}\xi_j^2 \delta\psi ^{\ast}_j(\bx,t) 
+gN\beta _{12}\xi^\ast_{\jbar}\xi_j\delta\psijbar \nonumber \\
&\qquad+gN\beta _{12}\xi_{\jbar} \xi_j \delta\psi ^{\ast}_{\jbar}(\bx, t)
\,.
\end{align}
Then, $\delta  \psi_j$ are expanded as
\begin{equation}
\delta\psij= 
\sum_q \left\{
u _{qj}(\bx )e^{-i E_q t}+v _{qj}^{\ast}(\bx )e^{i E_q^{\ast} t}
\right\} \,. \label{deltapsi}
\end{equation}
Here, $u _{qj}$ and $v _{qj}$ are eigenfunctions of the
following BdG equation,
\begin{equation}
\mathcal{T}\bu _q=E _q\bu _q \, , \label{BdG}
\end{equation}
where the quartet representation is introduced,
\begin{align} 
\bu_q&=
\begin{pmatrix}
u_{q1}\\
u_{q2}\\
v_{q1}\\
v_{q2} 
\end{pmatrix}
\,, \qquad\qquad
\mathcal{T}=
\begin{pmatrix}
\cal{L} &\cal{M} \\
-\cal{M}^{\ast} &-\cal{L}^{\ast}
\end{pmatrix}  
\,, \\
\mathcal{L}&=
\begin{pmatrix}
\mathcal{L} _{11} &\mathcal{L} _{12} \\
\mathcal{L} _{21} &\mathcal{L} _{22} \\
\end{pmatrix}
\,,\qquad \cal{M}=
\begin{pmatrix}
\mathcal{M} _{11} &\mathcal{M} _{12} \\
\mathcal{M} _{21} &\mathcal{M} _{22}
\end{pmatrix}\,,
\\
\mathcal{L} _{jj}&=h_0-\mu _j+2gN\beta _{jj}|\xi _j|^2+gN\beta _{12}|\xi _{\jbar}|^2\,, \\
\mathcal{M} _{jj}&=gN\beta _{jj}\xi _j^2\,, \\
\mathcal{L} _{12}&=\mathcal{L} _{21}^\ast = gN\beta _{12}\xi _2^\ast\xi _1\,, \\
\mathcal{M} _{12}&=\mathcal{M} _{21}=gN\beta _{12}\xi _2\xi _1 \,.
\end{align}
We define the indefinite inner product for 
any pair of quartets ${\bm s}(\bx)$ and
${\bm t}(\bx)$ by
\begin{align}
(\bm s, \bm t) =\int\! dxdy\, \bm s^\dagger(\bx)\tau_3{\bm t}(\bx) \,,\quad 
 \tau_i = \sigma_i \otimes 1 = 
\begin{pmatrix}
I &0 \\
0 &-I
\end{pmatrix} \,,
\end{align}
where $\sigma_i$ ($i=1,2,3$ ) are the Pauli matrices, operating on the space of
the doublet $(u_j,v_j)$, and $I$ is a $2\times2$ unit matrix with respect to the index $j$\,.
The symmetric property,
\begin{equation}
\tau_3 \mathcal{T} \tau_3= \mathcal{T}^\dagger\,, \label{symmetry3}
\end{equation}
leads to the pseudo-Hermiticity
of $\mathcal{T}$,
\begin{equation}
(\bm s,\mathcal{T} \bm t)= (\mathcal{T}\bm s, \bm t) \,.
\end{equation}
The squared norm of ${\bm s}(\bx)$,
\begin{equation}
\Vert \bm s \Vert ^2 =(\bm s,\bm s) \, ,
\end{equation}
can be positive, negative, and zero. Because the unperturbed eigenfunctions relevant to our discussion
belong solely to real eigenvalues, we do not repeat the properties of 
eigenfunctions belonging to complex and zero eigenvalues.
The symmetric property,
\begin{equation}
\tau_1 \mathcal{T} \tau_1=-\mathcal{T}^\ast\,,
\label{symmetry1}
\end{equation}
implies that, 
for each eigenfunction ${\by}_q$ ($\mathcal{T}{\by}_q=E_q {\by}_q$) 
belonging to a real eigenvalue
that is normalized as $\Vert \by_q \Vert ^2 =1$, there is an eigenfunction 
${\bz}_{\Tilde q}=\tau_1{\by}^\ast_q$ such that
$\mathcal{T} \bz_{\Tilde q}= -E_{q}\bz_{\Tilde q}$ with $\Vert \bz_{\Tilde q} \Vert ^2 =-1$\,.
Note that $\bz_{\Tilde q}$ may be denoted simply by $\bz_{q}$,
but we adopt the notation $\bz_{\Tilde q}$ to make our expressions simpler. 
The explicit form of ${\Tilde q}$ 
will be given below Eq.~(\ref{eq:z0}) in Sect.~\ref{sec-Application}.
The set of $\{\by_q\,,\, \bz_q\}$ is orthonormal,
\begin{equation}
(\by_q,\by_{q'}) = - (\bz_q,\bz_{q'})= \delta_{qq'} \,, \qquad
(\by_q,\bz_{q'})=0 \,,
\end{equation}
and complete,
\begin{align}
\sum _q \left[\by _q(\bx) \by ^{\dagger}_q(\bx' )-\bz _q(\bx) \bz ^{\dagger}_q(\bx' )\right] 
=\tau _3\delta(\bx -\bx' ) \label{complete}\,.
\end{align}

\section{GENERAL ANALYTIC FORMULATION BASED ON PERTURBATION THEORY}\label{sec-General}

The complex eigenvalue in the BdG equations~(\ref{BdG}) indicates 
the dynamical instability of the system.
In this study, we seek the parameter regions of the emergence of complex eigenmodes
following the analytical method in Ref.~\cite{Nakamura1},
which was originally applied to a single 
component system. We extend the work of Ref.~\cite{Nakamura1} to a two-component system
as follows. We suppose the vicinity of a boundary in the parameter space and divide it into regions with and without complex eigenvalues. 
We solve the stationary GP eq.~(\ref{stationaryGP}) and 
BdG eq.~(\ref{BdG}) to obtain their eigenvalues and eigenfunctions 
at a point belonging to the region without complex eigenvalues. These eigenvalues and eigenfunctions are regarded as
unperturbative eigenvalues and eigenfunctions. 
We then consider small variations in the parameters and develop a perturbative expansion
to find complex eigenvalues in the first order of the expansion 
when the parameter
variation crosses the boundary.

We develop the perturbative expansion as follows.
First, the quantities in the GP equation are expanded as
\begin{eqnarray}
\xij &=&\xi _j^{(0)}(\bm x ) +\varepsilon \xi _j^{(1)}(\bm x )+O(\varepsilon^2) \,,\\
\mu _j&=&\mu _j^{(0)}+\varepsilon\mu _j^{(1)} + O(\varepsilon^2) \,,
\end{eqnarray}
where $\varepsilon$ is an infinitesimal parameter that characterizes the parameter variation.
The expansion of the matrix $\mathcal{T}$, which involves both $\xi _j$ and $\mu_j$, 
is
\begin{equation}
\mathcal{T}=\mathcal{T}_0+\varepsilon \mathcal{T}'+O(\varepsilon^2) \,,
\end{equation}
where 
\begin{equation}
\mathcal{L} = \mathcal{L} _0+\varepsilon \mathcal{L} '+O(\varepsilon^2)\,, \quad
\mathcal{M} = \mathcal{M} _0+\varepsilon \mathcal{M} '+O(\varepsilon^2)\,,
\end{equation}
and
\begin{equation}
\mathcal{T} _0=
\begin{pmatrix}
\mathcal{L} _0 &\mathcal{M} _0 \\
-\mathcal{M} _0^{\ast} &-\mathcal{L} _0^{\ast} \\
\end{pmatrix} \,,
\qquad
\mathcal{T}'=
\begin{pmatrix}
\mathcal{L} ' &\mathcal{M} ' \\
-\mathcal{M} '^{\ast} &-\mathcal{L} '^{\ast} \\
\end{pmatrix} \,.\label{calTprime}
\end{equation}
Note that the symmetric properties (\ref{symmetry3}) and (\ref{symmetry1})
are respected in the perturbative expansion and that the properties
of the indefinite inner product are preserved at any order of the perturbation.

Likewise, the eigenfunctions and eigenvalues of the BdG equations are expanded as
$\bu _q^{(0)}+\varepsilon \bu _q^{(1)}+O(\varepsilon ^2)$ and
$ E_q=E_q^{(0)}+\varepsilon E_q^{(1)}+O(\varepsilon ^2)\,$, respectively.
The zeroth-order equations are
\begin{align}
\mathcal{T}_0\bu _q^{(0)}=E_q^{(0)}\bu _q^{(0)} \label{0BdG} \,,
\end{align}
and the first-order equations are organized as
\begin{align}
(\mathcal{T}_0-E_q^{(0)}) \bu _q^{(1)}=(E_q^{(1)}-\mathcal{T}')\bu _q^{(0)} \label{1BdG} \,.
\end{align}

Assuming that $E_q^{(0)}$ is real, we examine whether $E_q^{(1)}$ is complex.
According to Ref.~\cite{Nakamura1}, the necessary prerequisite to complex $E_q^{(1)}$
is a degeneracy between $\by^{(0)}_q$ and $\bz^{(0)}_{q}$\, but not between 
$\by^{(0)}_q$'s nor $\bz^{(0)}_q$'s. For a single $E_q^{(0)}$, consider a general situation in which 
there are $r$-fold degenerate
 $\by^{(0)}_{qi}$ $(i=1,2,\cdots,r)$
and $s$-fold degenerate $\bz^{(0)}_{q'i'}$ $(i'=1,2,\cdots,s)$\,,
{\it i.e.}, a total of $r+s$ degenerate states.
Here, $\bz^{(0)}_{q'}$ is the solution of
\begin{equation}
\mathcal{T}_0\bz _{q'}^{(0)}
\left(=-E_{\tilde q'}^{(0)}\bz _{q'}^{(0)}\right)
 =E_q^{(0)}\bz _{q'}^{(0)}   \,. 
\end{equation}
Then, $\bu _q^{(0)}$ is generally given by their linear combination,
\begin{equation}
\bu _q^{(0)} =\sum _{i=1}^r c_{y i}\by _{q i}^{(0)}
 +\sum _{i'=1}^s c_{z i'}\bz _{q' i'}^{(0)} \,.
\end{equation} 
Substituting this linear expression into Eq.~(\ref{1BdG}), we obtain the secular equation
for $E_q^{(1)}$\,,
\begin{align}
&
\begin{vmatrix}
\left( \by _{q 1}^{(0)},\mathcal{T}'\by _{q 1}^{(0)}\right)- E_q^{(1)}
 &\cdots &\left( \by _{q 1}^{(0)},\mathcal{T}'\bz _{{q'} 1}^{(0)}\right)
 &\cdots \\
\vdots & \ddots &\vdots & \ddots\\
\left( \bz _{{q'} 1}^{(0)},\mathcal{T}'\by _{q 1}^{(0)}\right)
 &\cdots & \left( \bz _{{q'} 1}^{(0)},\mathcal{T}'\bz _{{q'} 1}^{(0)}\right)+
 E_q^{(1)} &\cdots\\
\vdots & \ddots &\vdots & \ddots
\end{vmatrix} \nonumber \\
& =0 \,.
\end{align}
Multiplying each row  from the $(r+1)$-th row to
the $(r+s)$-th row by $-1$, we rewrite this secular equation as 
\begin{align}
\left| T'-E_q^{(1)} \right|=0 \,,
\label{CharaEq}
\end{align}
with
\begin{alignat}{2}\label{T}
T'&=\begin{pmatrix}L_y &M\\-M ^\dagger &-L_z \end{pmatrix} \,,\quad&
(L_y)_{ii'} &=\left(\by _{q i}^{(0)} ,\mathcal{T}'\by _{q i'}^{(0)}\right)\\
(L_z)_{ii'} &=\left(\bz _{{q'} i}^{(0)} ,\mathcal{T}'\bz _{{q'} i'}^{(0)}\right) \,,\quad&
 M_{ii'}&=\left(\by _{q i}^{(0)} ,\mathcal{T}'\bz _{{q'} i'}^{(0)}\right)
\,. \label{M}
\end{alignat}
It can be proven from the pseudo-Hermiticity of $\mathcal{T}'$
that $L_y^\dagger= L_y$ and $L_z^\dagger= L_z$. 
When $M$ does not vanish, $T'$ is non-Hermitian, and $E_q^{(1)}$ can be complex.

Our procedure for investigating the dynamical instability of a system
consists of the following four steps. (1) We find appropriate zeroth-order
BdG equations with real eigenvalues and solve the equations to obtain $\by^{(0)}$
and $\bz^{(0)}$\,. (2) The condition for the degeneracy between
$\by^{(0)}$ and $\bz^{(0)}$ is determined. (3) The secular equation involving
degenerate $\by^{(0)}_{qi}$ and $\bz^{(0)}_{{q'}i'}$ is established.
(4) We verify whether the first-order eigenvalue $E^{(1)}_{q}$ is complex or
real by solving the secular equation.

\section{APPLICATION TO TWO-COMPONENT QUANTIZED VORTICES}\label{sec-Application}

In this study, we consider a trapped two-component system with quantized vortices, 
characterized by winding numbers $\kappa_j$ for 
component $j$\, $(j=1,2)$\,. Both vortex cores are located at the center of the
trapping potential, which is set to the origin. We assume that all particle 
interactions, both intra- and inter-component interactions, are weak. That is, the coupling
constant $g$ in Eq.~(\ref{TDGP}) is a small parameter on the order of $\varepsilon$,
and a perturbation expansion with respect to $g$ is developed. For this purpose,
we replace $g$ with $\varepsilon g$\,. Then, the vortex
solutions of the stationary GP equations with $\kappa_j$ are
\begin{equation}
\xi _j(r ,\theta)=\sqrt{\frac{1}{2\pi}}e^{i\kappa _j\theta}f_j(r)\,,
\end{equation}
where $r$ and $\theta$ are the polar coordinates. Without loss of generality, the 
range of $\kappa_j$'s 
may be restricted to
\begin{equation}
\kone \geq 0 \,, \qquad  \kone \geq |\ktwo | \,.
\label{eq:kappa12}
\end{equation}
We call the rotations for $\ktwo\geq0$ and $\ktwo<0$ 
co- and counter-rotations, respectively.

\subsection{Zeroth-Order BdG Equations}\label{subsec-0BdG}

For $g=0$, the BdG equations~(\ref{0BdG}) are 
linear Schr\"odinger equations under the isotropic harmonic potential 
and can be solved analytically, 
irrespective of the stationary solutions of the GP equations.
All eigenvalues are real.

The zeroth-order eigenvalues and eigenfunctions of the BdG equations
are labeled by $q=(n,\ell,j)$; $n\,,\,\ell$ being
the principal and magnetic quantum numbers, respectively, and $j$ representing
the component index. 
To give the eigenfunctions of the BdG and GP equations, 
we introduce the eigenfunctions $\phi_{n\ell j}$\,,
\begin{equation}
\left\{h_0-\mu^{(0)}_j \right\} \phi_{n\ell j}=E^{(0)}_{n\ell j} \phi_{n\ell j}
\,,
\end{equation}
which are given by
\begin{align}
E^{(0)}_{n\ell j} &= \omega \left( 2n +|\ell+\kappa_j|+1\right)
-\mu_j^{(0)} \,\\
n&=0,1,2,\cdots \, , \qquad \ell = 0,\pm 1,\pm 2\,,\cdots \,,
\end{align}
and
\begin{equation}
\phi _{n\ell j}(\rho ,\theta )=\sqrt{\frac{1}{2\pi}}e^{i(\ell+\kappa _j)\theta}
R_{n\ell j}(\rho ) \label{phi}\,,
\end{equation}
with $\rho = \sqrt{m\omega} r$\,.  The explicit forms of $R_{n\ell j}(\rho ) $ are 
presented in Appendix~\ref{app-0BdG}. The zeroth-order BdG eigenfunctions are
\begin{alignat}{2}
\by _{n \ell 1}^{(0)}&=
\begin{pmatrix}
\phi_{n\ell 1} \\
0 \\
0 \\
0 
\end{pmatrix}
\,,\qquad&
\by _{n \ell 2}^{(0)}&=
\begin{pmatrix}
0 \\
\phi_{n\ell 2}\\
0 \\
0 
\end{pmatrix}\,,\\
\bz_{n \ell 1}^{(0)}&=
\begin{pmatrix}
0 \\
0 \\
\phi_{n -\ell 1}^\ast  \\
0 
\end{pmatrix}
\,,\qquad&
\bz _{n\ell 2}^{(0)}&=
\begin{pmatrix}
0 \\
0 \\
0 \\
\phi_{n -\ell 2}^\ast
\end{pmatrix}\,.
\label{eq:z0}
\end{alignat}
Note the definition $\Tilde{q} = (n,-\ell, j)$ for $q = (n,\ell, j)$, which implies 
that $\bz_{n\ell j} = \tau_1 \by_{n-\ell j}^*$ and that 
the eigenvalues of $\by_{n\ell j}$ and $\bz_{n\ell j}$ are $E^{(0)}_{n\ell j}$ and 
$E^{(0)}_{n-\ell j}$, respectively. 
We normalize $\phi_{n\ell j}$,
\begin{equation}
\int_0^\infty \int_0^{2\pi}\left|\phi_{n\ell j}(\rho,\theta) \right|^2\, 
rdr d\theta =1\,,
\label{phinormal}
\end{equation}
so
\begin{equation}
(\by_{n\ell j}, \by_{n'\ell' j'})=-(\bz_{n\ell j}, \bz_{n'\ell' j'})= \delta_{nn'}\delta_{\ell\ell'}\delta_{jj'}\,.
\end{equation}

\subsection{Zeroth- and First-Order GP Equations}

The zeroth-order stationary GP equations are
\begin{equation}
h_0 \xi _{j}^{(0)}=\mu _j^{(0)}\xi _j^{(0)}
\,.
\end{equation}
Their solutions, which are the lowest eigenstates, are found to be
\begin{equation}
\xi_ j^{(0)}(\rho,\theta )=\phi_{00 j}(\rho,\theta ) \,,\qquad \mu _j^{(0)}
=\omega\left(|\kappa _j|+1\right) \,.
\end{equation}

Next, we have the first-order stationary GP equations,
\begin{align}
& \left\{h_0-\mu _j^{(0)}\right\}\xi ^{(1)}_j \nonumber \\
& =
-\left( gN \beta _{jj}\left|\xi ^{(0)}_j \right|^{2}+ gN\beta _{12}\left|\xi ^{(0)}_{\jbar} 
\right|^{2}-\mu _j^{(1)}\right )\xi ^{(0)}_j \,.
\end{align}
Multiplying both sides by $\xi_j^{(0)\ast}$ and integrating them over the whole 
two-dimensional space, we obtain the first-order chemical potentials $\mu _j^{(1)}$  as
\begin{align}
\mu _j^{(1)}=m\omega F
\left( \frac{\beta _{jj}}{2^{2|\kj |}}\frac{(2|\kj| )!}{(|\kj| !)^2}+\frac{\beta _{12}}{2^{(\kone +|\ktwo |)}}\frac{(\kone +|\ktwo |)!}{\kone !|\ktwo |!} \right) \,,
\end{align}
with $F = {gN}/{2\pi} \,.$

\subsection{First-Order Matrix Elements}
The $\theta$-dependence of the first-order matrix $\mathcal{T}'$
 in Eq.~(\ref{calTprime}) can be factorized as
\begin{align}
\mathcal{T}'=
\begin{pmatrix}
\mathcal{U} &0 \\
0 &\mathcal{U}^\dagger \\
\end{pmatrix}
\begin{pmatrix}
\mathcal{L}'_r &\mathcal{M} '_r \\
-\mathcal{M} '_r &-\mathcal{L} '_r \\
\end{pmatrix} 
\begin{pmatrix}
\mathcal{U}^\dagger &0 \\
0 &\mathcal{U} \\
\end{pmatrix}
\,, \label{calTprimeUprime}
\end{align}
where the $r$-dependent matrix is
\begin{align}
\mathcal{L}'_r&=
-
\begin{pmatrix}
\mu^{(1)}_1 &0 \\
0 &\mu^{(1)}_2 \\
\end{pmatrix}
+2 F
\begin{pmatrix}
\beta_{11} R_{001}^2 &0 \\
0 &\beta_{22}R_{002}^2\\
\end{pmatrix}
\nonumber \\
&\hspace{20pt}+F \beta_{12}
\begin{pmatrix}
R_{002}^2 &R_{001} R_{002} \\
R_{001} R_{002} &  R_{001}^2\\
\end{pmatrix} \,, \\
\mathcal{M}'_r&=
F 
\begin{pmatrix}
\beta_{11} R_{001}^2 &\beta_{12}R_{001} R_{002} \\
\beta_{12} R_{001} R_{002}&\beta_{22} R_{002}^2\\
\end{pmatrix} \,, 
\end{align}
and the $\theta$-dependent unitary matrix $\mathcal{U}$ is
\begin{align}
\mathcal{U}=
\begin{pmatrix}
e^{i\kappa_1 \theta} &0 \\
0 &e^{i\kappa_2 \theta} \\
\end{pmatrix} \,.
\end{align}
It follows from Eq.~(\ref{calTprimeUprime}) that 
all phase factors $e^{i \kappa_j\theta}$ in the integrands of
the matrix elements~(\ref{T})--(\ref{M}) are canceled out and 
that only phase factors $e^{i \ell \theta}$ from $\by _{q, i}^{(0)}$ and
$\bz _{{q'}, i'}^{(0)}$ survive. Therefore, after  $\theta$-integration, the matrix elements 
carry $\delta_{\ell\ell'}$.
The matrix elements are evaluated as follows:
\begin{align}
&\phantom{=}\left(\by _{n\ell j}^{(0)},{\cal T}'\by _{n'\ell ' j}^{(0)}\right)=
\left(\bz _{n{-}\ell j}^{(0)}, {\cal T}'\bz _{n'{-}\ell ' j}^{(0)}\right)^\ast \nonumber\\
&=\delta_{\ell\ell'} \left[ 
-\delta _{nn'}\mu _j^{(1)} \right. +2F \beta _{jj}\int\! rdr \left\{R_{00j}^2R_{n\ell j} R_{n'\ell j}\right\}
 \nonumber \\ 
&\phantom{=\left(\by _{n\ell j}^{(0)}, {\cal T}'\by _{n'\ell ' j}^{(0)}\right)} \left. +F\beta _{12}\int\! rdr\left\{R_{00\jbar}^2
R_{n\ell j} R_{n'\ell j} \right\} \right] \label{matrix_yy1} 
\,,\\[10pt]
&\phantom{=}\left(\by _{n\ell j}^{(0)}, {\cal T}'\by _{n'\ell ' \jbar}^{(0)}\right)=
\left(\bz _{n{-}\ell j}^{(0)},{\cal T}'\bz _{n'{-}\ell ' \jbar}^{(0)}\right)^\ast \nonumber\\
&=\delta_{\ell\ell'} F\beta _{12}
\int\! rdr \left\{R_{00j}R_{00\jbar}R_{n\ell j} R_{n'\ell \jbar}\right\}
\,,\\[10pt]
&\phantom{=}\left(\by _{n\ell j}^{(0)}, {\cal T}'\bz _{n' {-}\ell ' j}^{(0)}\right)=
\left(\bz _{n'{-}\ell' j}^{(0)}, {\cal T}' {\by} _{n\ell  j}^{(0)}\right)^\ast \nonumber\\
&=\delta_{\ell{-}\ell'} F\beta _{jj} 
\int\! rdr \left\{R_{00j}^2R_{n\ell j} R_{n'-\ell {j}}\right\}
\,,\\[10pt]
&\phantom{=}\left(\by _{n\ell j}^{(0)}, {\cal T}'\bz _{n' {-}\ell ' \jbar}^{(0)}\right)=
\left(\bz _{n' {-}\ell' \jbar}^{(0)}, {\cal T}' {\by} _{n\ell  j}^{(0)}\right)^\ast \nonumber\\
&=\delta_{\ell {-}\ell'} F\beta _{12}
\int\! rdr \left\{R_{00j}R_{00\jbar}R_{n\ell j} R_{n'-\ell \jbar}\right\}
\,.
\end{align}

The above properties of the matrix elements allow matrix $T'$ in Eq.~(\ref{T}) to be shifted to a block diagonal form. 
We classify the degenerate states $\by_q$'s and $\bz_{q'}$'s
into groups according to the value of $\ell$, that is, $m_f$ groups labeled by $\ell =\ell_1,\cdots, \ell_{m_f}$\,.
Rearranging the matrix elements according to the above groups, we obtain the following block
diagonal matrix $T'$: 
\begin{align}
T' =
\begin{pmatrix}
T'_{\ell _1} &0 &0 &0 \\
0 &T'_ {\ell _2} &0 &0 \\
0&0 & \ddots &0 \\
0 &0 &0 &T'_ {\ell_{m_f}}\\
\end{pmatrix} \,.
\label{TprimeBlock}
\end{align}

\subsection{Patterns of Degeneracy}
Based on the conclusions of the previous subsection, complex $E^{(1)}$ can only appear in block matrix $T'_{\ell_m}$\, that
includes both $\by_q$ and $\bz_{q'}$. In addition, the matrix $M_{\ell_m}$ in $T_{\ell_m}$ must be 
non-vanishing. We, therefore, seek conditions for non-vanishing 
$M_{\ell}=\left(\by _{n\ell j}^{(0)}, \mathcal{T}'\bz _{n'\ell j'}^{(0)}\right)$\,.
Hereafter, the superscript ${}^{(0)}$ for $\by$ and $\bz$ is implicit for the sake of simplicity. 
The eigenvalues of $\by_{n\ell j}$ and $\bz_{n'\ell j'}$ are
\begin{align}
E_{n\ell j}^{(0)}&=\omega \left( 2n +|\ell+\kappa _j|-|\kappa _j|\right) \\
-E_{n'-\ell j'}^{(0)}&=- \omega \left( 2n' +|-\ell+\kappa _{j'}|-|\kappa _{j'}|\right)
\end{align} 
The degeneracy condition between $\by_{n\ell j}$ and $\bz_{n'\ell j'}$, 
namely $E_{n\ell j}^{(0)}=-E_{n'-\ell j'}^{(0)}$\, is
\begin{equation}
2(n+n')+|\ell+\kappa _j|+|\ell-\kappa _{j'}|-|\kappa _j|-|\kappa _{j'} | =0\,.
\label{DegeneCondYZ}
\end{equation}
When complex $E^{(1)}_{q}$ is found, it can be shown from Eq.~(\ref{symmetry3}) that $E^{(1)}_{\Tilde q}$ is also complex. 
Therefore, without loss of generality, $\ell$ can be restricted to $\ell\geq 0$ when searching for the condition for complex eigenvalues. The degeneracy is
possible only when $E_{n\ell j}^{(0)} \leq 0$ and/or $-E_{n'-\ell j'}^{(0)}\geq 0$,
which restricts the allowed value of $\ell$ to $0\leq \ell \leq -2\kappa_j$ for
 $\kappa_j \leq 0$ and 
$0\leq \ell \leq 2\kappa_{j'}$ for $ \kappa_{j'} \geq 0$\,. Finally, with 
Eq.~(\ref{eq:kappa12}),
we only have to consider the range $0\leq \ell \leq 2 \kappa_1$\,.
The solutions of Eq.~(\ref{DegeneCondYZ}) are categorized 
into the following four types of ($j,j'$) : 
\def\labelenumeii{(\arabic{enumii})}
\begin{enumerate}
	\item[(a)] $(j,j') = (1,1)$: $n=n'=0$ when $0<\ell \leq \kappa_1$
	\item[(b)] $(j,j') = (2,2)$: $n=n'=0$ when $0<\ell \leq |\kappa_2|$
	\item[(c)] $(j,j') = (1,2)$: $n=n'=0$ when $0 < \ell \leq \kappa_2$
	\item[(d)] $(j,j') = (2,1)$: \vspace{-5pt}
	\begin{itemize}
		\item $n=n'=0$ \hspace{3mm} when $0<\ell \leq \kappa_1$ and $0 \leq \kappa_2$
		\item $n+n'=-\kappa_2$ when $0 \leq -\kappa_2 \leq \ell \leq \kappa_1$
		\item $n+n'=\ell$ \hspace{3.6mm} when $0 <\ell\leq -\kappa_2$
		\item $n+n'=-\ell + \kappa_1-\kappa_2$ when $\kappa_1 \leq \ell \leq \kappa_1 - \kappa_2$
	\end{itemize}
\end{enumerate}
\begin{figure}[tbh!]
	\centering
	\includegraphics[width=8.6cm]{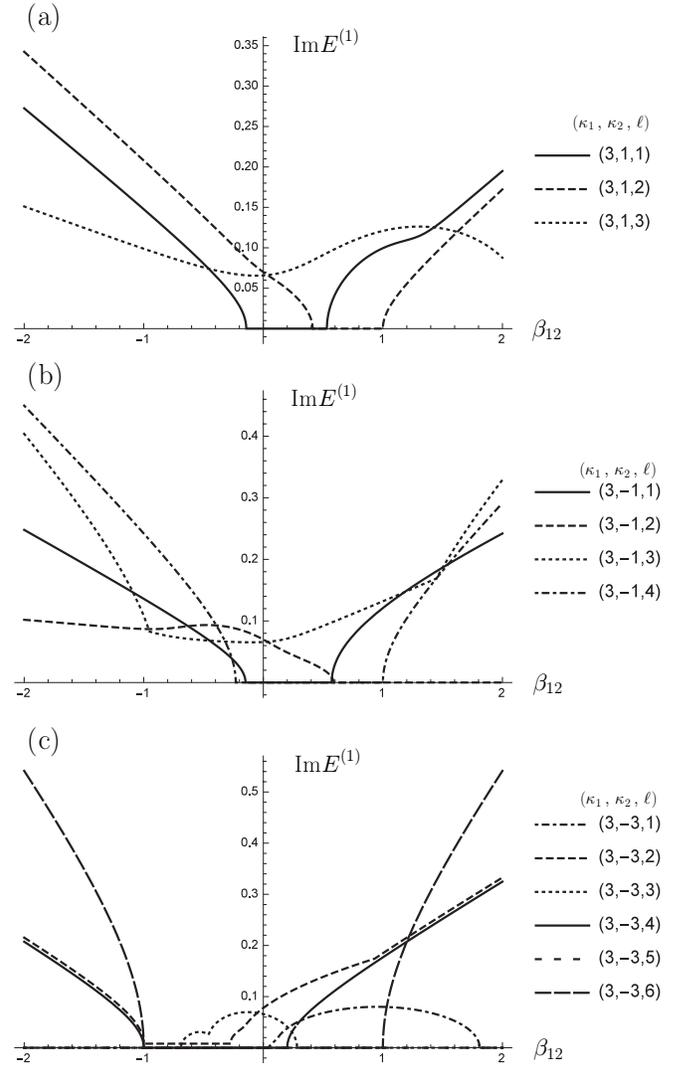}
	\caption{Regions with possible double degeneracy between 
		$\by_{n\ell j}$ and $\bz_{n'\ell j'}$ in the $\kappa_2-\ell$ plane
		for a fixed $\kappa_1 >0$. 
		They are 
		(a) $\by_{0\ell1}$ and $\bz_{0\ell1}$, 
		(b) $\by_{0\ell2}$ and $\bz_{0\ell2}$, 
		(c) $\by_{0\ell1}$ and $\bz_{0\ell2}$, 
		(d) $\by_{n\ell2}$ and $\bz_{n'\ell1}$. 
		 The number inside each subregion denotes 
		the value of $n+n'$.}
	\label{fig:regions}
\end{figure}
Note that the modes with $(n,\ell,j) = (0,0,1)\,,\,(0,0,2)$ 
are excluded from our considerations because they are zero modes. 
These modes remain as zero modes and never turn into complex modes under perturbation that 
retains the global phase symmetries $\xi_j \to \xi_j e^{i\delta_j}$ \cite{Takahashi2014}.
From the inequality,
\begin{align}
E_{0\ell j}^{(0)}+E_{0-\ell j'}^{(0)}=
|\ell+\kappa _j|+|\ell-\kappa _{j'}|-|\kappa _j|-|\kappa _{j'}| \geq 0
\end{align}
for $(j,j')= (1,1),(2,2),(1,2)$ and $(j,j')=(2,1)$ with $\kappa_2\geq 0$,
we see that radial excited states $n+n' \neq 0$ participate only in (d) with 
$\kappa_2 < 0$\, implying counter-rotation.

\begin{figure}[tbh!]
	\centering
	\includegraphics[width=6cm]{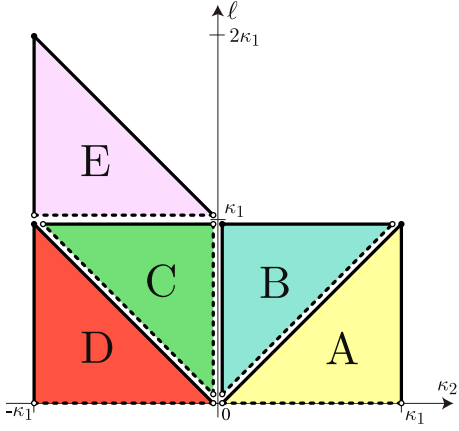}
	\caption{Subregions with possible multiple degeneracies involving both
		$\by$ and $\bz$. The boundaries denoted by solid lines or filled circles
		are included in the subregion 
		while those denoted by dashed lines or open circles are excluded from the subregion.}
	\label{fig:regions2}
\end{figure}

The regions with double degeneracy
in the $\kappa_2-\ell$ plane for a fixed $\kappa_1 >0$
are depicted in Fig.~\ref{fig:regions}. Collecting all these results,
we obtain all possible multiply degenerate sets
that contain both $\by$ and $\bz$. 
They are categorized into the following five subregions: 
[A] $0 < \ell \leq \kappa_2$,\,
[B] $0 \leq \kappa_2 < \ell \leq \kappa_1$,\,
[C] $0 < -\kappa_2 < \ell \leq \kappa_1$,\,
[D] $0 < \ell \leq -\kappa_2$,\,
[E] $\kappa_1 < \ell \leq \kappa_1-\kappa_2$\, 
in Fig.~\ref{fig:regions2}.
For example, in  
subregion C, where regions (a) and (d) with double degeneracy
overlap one another but do not overlap regions (b) nor (c), 
we find two types of degenerate sets, namely
($\by_{0 \ell 1}$,$\by_{-\kappa_2 \ell 2}$,$\bz_{0 \ell 1}$) and
 ($\by_{n \ell 2}$,$\bz_{n'\ell 1}$), where $n+n'=-\kappa_2$ and $n\ne-\kappa_2$.
All types of degenerate sets are summarized in Table~\ref{table:dmembers}. 
The types are labeled by numbers representing degrees of 
degeneracy and by the additional indices $y, z,$ and $n$ in C, D and E. 

\begin{figure}[tbh!]
	\centering
	\includegraphics[width=8.3cm]{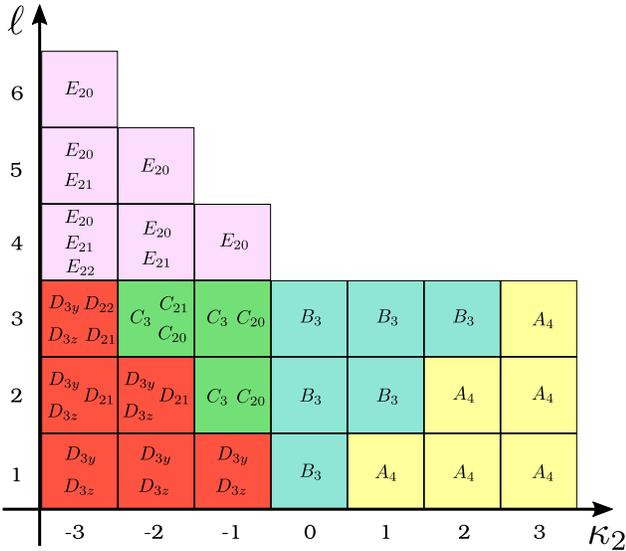}
	\caption{(Color online) All degenerate types for $\kappa_1=3$}
	\label{fig:Degeneracykappa13}
\end{figure}

For illustration, we count all degenerate sets for $\kappa_1=3$ 
in Fig.~\ref{fig:Degeneracykappa13}. Figure \ref{fig:Degeneracykappa13} and Table~\ref{table:dmembers}
provide general features of the appearance of the degenerate sets. 
In co-rotation, {\it i.e.}, for $\kappa_2 \geq 0$, there is only one degenerate set for each pair of $(\kappa_2, \ell)$, 
which are either A$_4$ or B$_3$, and no radially excited state is involved.
The degenerate patterns are richer in counter-rotations. 
The degenerate region extends to the maximum value $\ell =2\kappa_1$, 
which has been restricted to $\ell \leq \kappa_1$ for the co-rotating case. 
All degenerate sets involve radially excited states except for those on the line $\ell=-\kappa_2+ \kappa_1$.
There are plural degenerate sets for each $(\kappa_2, \ell)$. In particular, there are $1-\kappa_2$ sets in subregion C, 
$1+\ell$ in D, and $1+\kappa_1-\kappa_2-\ell$ in E.
These numbers increase around $\ell=\kappa_1$, where $-E_{0-\kappa_11}$ is positive and reaches a maximum, 
and as $\kappa_2$ approaches $-\kappa_1$.
The total number of degenerate sets for the winding number pair
 $(\kappa_1, \kappa_2)$, namely for that in each column in Table~\ref{table:dmembers},
 is $\kappa_1 - \kappa_2(\kappa_1+1)$ for the counter-rotating case 
 and $\kappa_1$ for the co-rotating case including $\kappa_2=0$.

\begin{table}[tbh!]
\begin{ruledtabular}
	\caption{All types of possible degeneracies involving both
		$\by$ and $\bz$ in each subregion and the relevant eigenfunctions.
		}
\begin{tabular}{ccl}\label{table:dmembers}
	Subreg. & Symbol& Degenerate set of eigenfunctions
	\\\hline
	A & A$_{4}$  & ($\by_{0 \ell 1}$, $\by_{0 \ell 2}$, $\bz_{0 \ell 1}$, $\bz_{0 \ell 2}$)\\
	B & B$_{3}$  & ($\by_{0 \ell 1}$, $\by_{0 \ell 2}$, $\bz_{0 \ell 1}$) \\
	C & C$_{3}$ & ($\by_{0 \ell 1}$, $\by_{-\kappa_2 \ell 2}$, $\bz_{0 \ell 1}$) \\
	  & C$_{2n}$ & ($\by_{n \ell 2}$, $\bz_{n'\ell 1}$)\; with $n+n'=-\kappa_2$; $n\ne-\kappa_2$ \\
	D & D$_{3y}$ & ($\by_{0 \ell 1}$, $\by_{\ell \ell 2}$, $\bz_{0 \ell 1}$) \\
	  & D$_{3z}$ & ($\by_{0 \ell 2}$, $\bz_{\ell \ell 1}$, $\bz_{0 \ell 2}$) \\
	  & D$_{2n}$ & ($\by_{n \ell 2}$, $\bz_{n' \ell 1}$)\; with $n+n'=\ell$; $n,n'\neq 0,\ell$ \\
	E & E$_{2n}$ & ($\by_{n \ell 2}$, $\bz_{n'\ell 1}$)\;  with $n+n'=\kappa_1-\kappa_2-\ell$
\end{tabular}
\end{ruledtabular}
\end{table}

\subsection{Range of Inter-Component Interaction Parameter $\beta_{12}$ of Complex Eigenvalues}

\begin{figure}[tb!]
	\centering
	\includegraphics[width=0.45\textwidth]{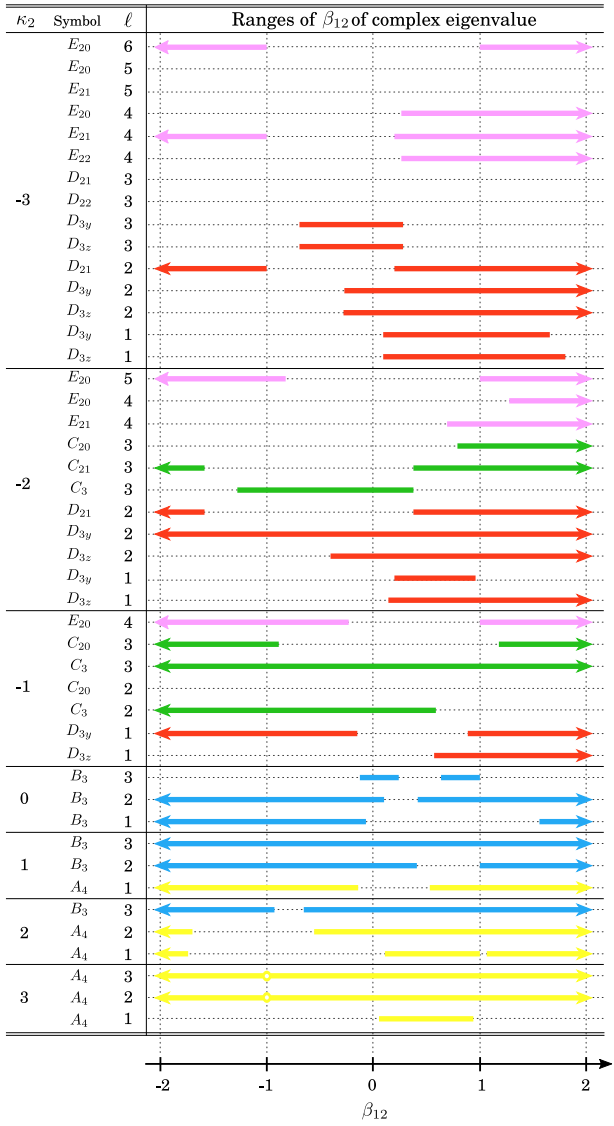}
	\caption{(Color online) Range of $\beta_{12}$ for complex eigenvalues in each $(\kappa_1=3,\kappa_2,\ell)$.}
	\label{fig:ComplexRegion}
\end{figure}

\begin{figure}[tb!]
	\centering
	\includegraphics[width=7.5cm]{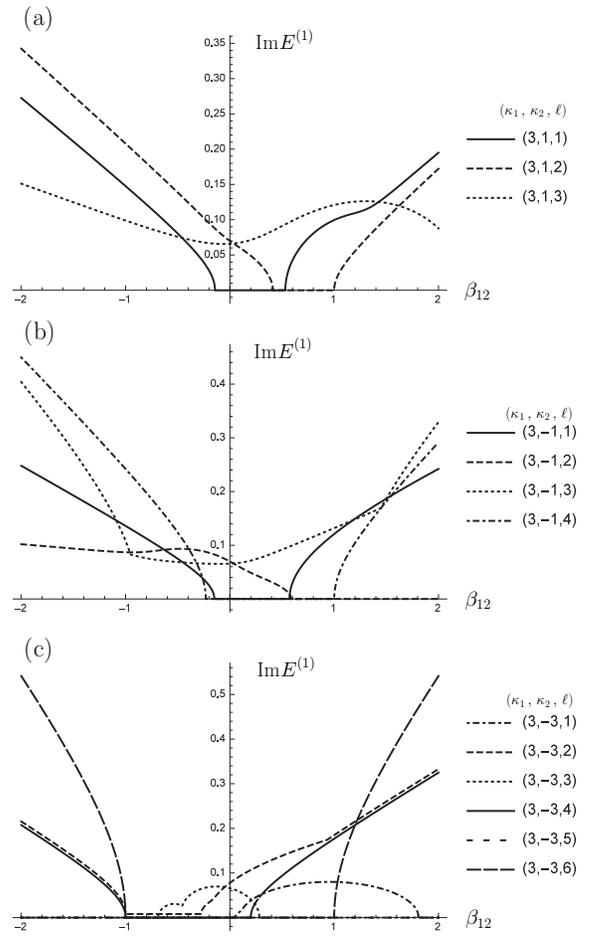}
	\caption{Maximum value of the imaginary parts of $E^{(1)}$ 
	in units of $m\omega F$
	for (a) $(\kappa_1,\kappa_2)=(3,1)$ with $\ell =1\, \sim \, 3$\,, 
	(b) $(\kappa_1,\kappa_2)=(3,-1)$ with $\ell =1\, \sim \, 4$, and
	(c) $(\kappa_1,\kappa_2)=(3,-3)$ with $\ell =1\, \sim \, 6$\,.}
	\label{fig:g3}
\end{figure}

First-order complex eigenvalues $E_{q}^{(1)}$ emerge only 
in subregions A--E, as shown above. Note that this is a prerequisite 
condition for the emergence of complex eigenvalues, and we have to verify
the secular equation Eq.~(\ref{CharaEq}) to determine whether its solution is complex
or real. We then study how the inter-component interaction affects
the stability of a system with two vortices.
Varying the inter-component interaction 
parameter $\beta_{12}$ with fixed $\beta_{11}=\beta_{22}=1$ for $\kappa_1=3$, we 
seek the ranges of $\beta_{12}$ in which some eigenvalues are complex. To this end, we 
manipulate only algebraic equations, which gives our study a clear advantage 
in numerical calculations over those that require solving the differential equations.  
Moreover, we use the discriminant $\Delta$ of the polynomial in the secure equation
for doubly and triply degenerates sets and obtain the range of $\beta_{12}$ for $\Delta$.
For quadruply degenerate sets $A_4$, we directly solve the secure equation and find
the ranges of the complex eigenvalues. The results are shown for each
$(\kappa_1,\kappa_2,\ell)$ in Fig.~\ref{fig:ComplexRegion}.
We also plot the maximum value of the imaginary parts of $E^{(1)}$ for 
$\kappa_1=3$, $\kappa_2=1,-1,-3$, and all possible $\ell$\, in Fig.~\ref{fig:g3}.


We can identify some general conclusions from Fig.~\ref{fig:ComplexRegion}.
The co-rotating systems tend to be dynamically unstable.
Most of the degenerate sets have wide ranges of $\beta_{12}$ with complex eigenvalues. 
The behaviors of counter-rotating systems of the two vortices are more complicated
because the number of degenerate sets at each $(\kappa_1,\kappa_2,\ell)$ is two or more.
For doubly degenerate sets, {\it i.e.}
C$_{2n}$, D$_{2n}$, and $E_{2n}$ in Fig.~\ref{fig:ComplexRegion}, 
some complex eigenvalues appear
away from $\beta_{12}=0$, while only real eigenvalues appear
for some $(\kappa_1,\kappa_2,\ell)$\,. Thus, the sign of $\beta_{12}$ is not 
essential, but positive $\beta_{12}$ is more likely to result in complex eigenvalues.
Ranges without complex eigenvalue exist in
 subregion $E$ over small $|\beta_{12}|$\,. 
This fact is consistent with the interpretation that decays of the two 
counter-rotating vortices are accelerated by energy exchange between the vortices
through inter-component interaction.
On the other hand, we also find stabilization due to the inter-component interaction. 
A weakly interacting system of a single vortex with winding number $\kappa$, 
which corresponds to the limiting case of $\beta_{12}=0$ in our formulation, 
gives complex eigenvalues for some $\ell$ 
and is dynamically unstable when $|\kappa|  \geq 2$ \cite{Nakamura1,Pu}.
It is remarkable that
there are a few ranges of $\beta_{12}$ where no complex eigenvalues arise (Fig.~\ref{fig:ComplexRegion}), 
explicitly $0.24 \leq \beta_{12} \leq 0.42$ for $\kappa_2=0$ 
and $-1.0 \leq \beta_{12} \leq -0.69$ for $\kappa_2=-3$.

Note that some of our results for $\kappa_2=-\kappa_1$ 
differ from those in Ref.~\cite{Ishino}, which were obtained by numerically solving the differential equations. However, the results are not inconsistent because the coupling constant $g$ 
in Ref.~\cite{Ishino} is much larger than that in our perturbative method.

\subsection{Splitting Patterns of Vortices}
Let us next consider the splitting patterns of vortices according to the method of Ref.~\cite{Kawaguchi}.

Because the zeroth-order degeneracy of interest is quartic at the highest,
we may express the zeroth-order eigenfunction with a fixed $\ell$, which may involve complex eigenvalues at the first order as
\begin{equation}
\bu_\ell = 
c_{y1} \by_{n_1  \ell 1} +  
c_{z1} \bz_{n_1' \ell 1} +
c_{y2} \by_{n_2  \ell 2} +
c_{z2} \bz_{n_2' \ell 2} \,,
\end{equation}
where the $c$ coefficients are determined from the eigenequation for
$T'_\ell$. One or two of $c$ coefficients are zero
for triple or double degeneracy in the subregions B--E.
When a mode associated with the eigenfunction $\bu_\ell$ is excited, 
the change in the $j$-component order parameter is, for example,
\begin{equation}
	\delta\psi_{\ell j}  \propto c_{yj} \phi_{n_j \ell j} +  c_{zj} \phi_{n_j -\ell j}^{*} \,,
\end{equation}
which grows exponentially in time. Then, the condensate density of the $j$-component deforms as
\begin{equation}
	|\psi_j + \delta\psi_{\ell j}|^2 \simeq  |\psi_j|^2 + 2 \mathrm{Re} \psi^*_j
 \delta\psi_{\ell j} \,,
\end{equation}
which has an $\ell$-fold rotational symmetry. 

\begin{figure}[tb]
	\centering
	\includegraphics[width=8.6cm]{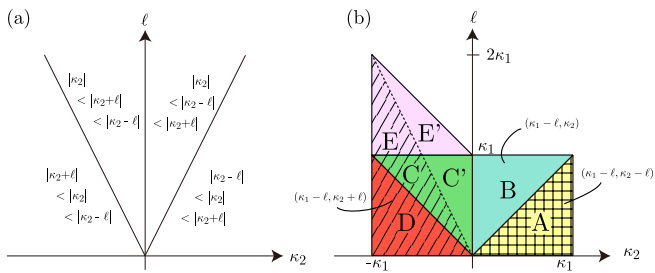}
	\caption{(Color online) (a) Magnitude relationship among $|\kappa_2-\ell|$, $|\kappa_2+\ell|$, 
		and $|\kappa_2|$. The region is divided into four areas by
		the lines $\kappa_2=0$ and $\ell = \pm 2\kappa_2$.
		 (b) Diagram of the splitting patterns of the vortices. 
		The pattern of the resultant winding numbers after splitting of the initial 
		$(\kappa_1, \kappa_2)$ are classified into three shaded areas.
	}
	\label{fig:magnitudeRelation}
\end{figure}
The asymptotic form of the $j$th-component order parameter in the limit of 
$\rho\to0$ is controlled by the winding number $\kappa_j$ as $\xi_j \sim  \rho^{|\kappa_j|}$ \,.
Inversely, the exponent of the asymptotic form of the order parameter informs the 
winding number. Because the asymptotic form of 
$\phi_{n\ell j}$ is proportional to $\rho ^{|\kappa_j+\ell|}$,
the winding number of the $j$th-component vortex 
with exponential growth results in 
$|\kappa_j\pm \ell|$ or remains $|\kappa_j|$. 
This restriction on the change in $\kappa_j$ is crucial. For $j=1$, we always have 
$|\kappa_1-\ell| < |\kappa_1| < |\kappa_1+\ell|$.
The restriction for $j=2$ is depicted in 
Fig.~\ref{fig:magnitudeRelation} (a). 
Combining these results with Fig.~\ref{fig:regions2}, we finally 
obtain the splitting diagram shown in Fig.~\ref{fig:magnitudeRelation} (b). 
For example, in subregion A, it is predicted that the winding numbers of
the two vortices vary from
$(\kappa_1, \kappa_2) \to (\kappa_1-\ell, \kappa_2-\ell)$ once complex eigenvalues arise because $|\kappa_j - \ell|$ are the smallest in A. At first glance, we may conclude that the winding numbers also 
change into $(\kappa_1-\ell, \kappa_2-\ell)$ in subregion B, but this is not true. The correct answer is 
$(\kappa_1-\ell, \kappa_2)$ because $\bz_{n\ell 2}$ is not a member of the 
triply degenerate set in subregion B [see Table \ref{table:dmembers}].
Subregions C and E are divided into two respective areas by the line $\ell=-2\kappa_2$.

\section{Summary}\label{sec-Summary}
In this study, we searched for the dynamical instability parameter regions of a two-component system 
with coaxial quantized vortices. Our analytical method applied perturbation with respect to the coupling constants.
Without numerically solving the BdG or TDGP equations, we completely obtained the 
unstable parameter ranges under the restriction of small coupling constants.
Our method consists of the following three steps. First, we list
all double degeneracies between $\by_{n\ell j}$ and $\bz_{n'\ell j'}$ 
at the unperturbed level, which is necessary for the emergence of complex eigenvalues 
at the first order of the perturbation. At this step, the unstable modes certainly satisfy
$1 \leq \ell \leq \kappa_1$ for the co-rotating system and $1 \leq \ell \leq \kappa_1-\kappa_2$ for the counter-rotating system.
Next, all multiple degeneracies 
involving both $\by$ and $\bz$ are enumerated. The relevant region of the $\kappa_2$--$\ell$ plane 
for a fixed $\kappa_1$ is divided into the
five subregions A--E, as shown in Table~\ref{table:dmembers}. Note that a variety of degeneracies appear in the two-component system; however, in the single-component system, only 
a double degeneracy is involved for each $\ell$, but no radial excitation (no $n \neq$ one) appears. 
Finally, we can determine whether the degeneracies raise complex eigenvalues  
by solving the secular equation within each candidate degenerate set. Because the inter-component interaction 
coupling constant $\beta_{12}$ is the most sensitive and interesting parameter, we have searched for the range of complex eigenvalues, and the results for $\kappa_1=3$ are shown in Figs.~\ref{fig:ComplexRegion} and \ref{fig:g3}. 
There are no heavy numerical calculations required because our secure equation 
is not more than a quartic equation. Thus, even though the system is restricted to small coupling constants, we have swept a wide parameter space, finding all unstable ranges.

As expected, Fig.~\ref{fig:ComplexRegion} shows that 
the co-rotating ($\kappa_2>0$) and counter-rotating ($\kappa_2 <0$) 
systems produce distinctive unstable ranges. In the co-rotating case, the addition of the 
second vortex ($\kappa_2$) does not drastically change the situation in which 
the highly quantized single vortex $\kappa_1\geq 2$ is already unstable. On the contrary, 
the counter-rotating system shows complicated behavior because the relative velocity 
of the two superflows are so large that large excitation energy radial modes 
can be members of degenerate sets, and complex
eigenvalues appear if energy exchange between the two fluid components is possible. 
As shown in Fig.~\ref{fig:ComplexRegion}, complex eigenvalues appear for $\ell$, which are 
larger than the winding numbers ( $\ell > \kappa_1\geq |\kappa_2|$), in subregion E. This tendency
becomes more pronounced for
 the larger absolute value of the inter-component coupling $|\beta_{12}|$,
which accelerates the energy exchange between two components.
We have also found a few regions where inter-component coupling stabilized the system. Finally, we have estimated all possible splitting patterns of the vortices with the aid of the degenerate set and the asymptotic forms of their eigenfunctions.

\begin{acknowledgments}
This work is supported in part by a Grant-in-Aid for Scientific Research (C) (No. 25400410) from the Japan Society for the 
Promotion of Science, Japan. 
\end{acknowledgments}
\appendix
\section{Function Related to the Zeroth-Order BdG Eigenfunction}
\label{app-0BdG}

We give a concrete expression of the function $R_{n\ell j}$\,,
from Eq.~(\ref{phi}) of Subsect.~\ref{subsec-0BdG},
\begin{align}
R_{n\ell j}(\rho )=
C_{n\ell j}e^{-\frac{1}{2}\rho ^2}\rho ^{|\ell+\kappa _j|}
S^{|\ell+\kappa _j|}_n(\rho ^2) \, .
\end{align}
Here, $S ^\gamma_n(x)$ is Sonine polynomial, defined by
\begin{eqnarray}
S ^\gamma _n(x)=\sum _{k=0}^n \frac{(-1)^k(n+\gamma )!}{(n-k)!(\gamma +k)!k!}x^k
\end{eqnarray}
with the orthonormal property,
\begin{equation}
\int ^{\infty}_0 S^{\alpha}_m(x)S^{\alpha}_n(x) x^{\alpha}e^{-x}\, dx 
=\frac{\Gamma(\alpha +n+1)}{n!}\delta _{mn} \,.
\label{SonineOrthoNormal}
\end{equation}
The normalization condition on $\phi_{n\ell j}$, Eq.~(\ref{phinormal}),
implying
\begin{align}
\int_0^\infty \left|R_{n\ell j}(\rho) \right|^2\, 
r dr=1 \,,
\end{align}
fixes the normalization factor,
\begin{equation}
C_{n\ell j} =\sqrt{\frac{2m\omega \, n!}{(n+|\ell+\kappa_j|)!}}\,.
\end{equation}


\begin{thebibliography}{99}

\bibitem{Shin}
Y.~Shin, M.~Saba, M.~Vengalattore, T.~A.~Pasquini, C.~Sanner,
A.~E.~Leanhardt, M.~Prentiss, D.~E.~Pritchard, and W.~Ketterle,
Phys.~Rev.~Lett. {\bf 93}, 160406 (2004).

\bibitem{Fallani}
L.~Fallani, L.~De Sarlo, J.~E.~Lye, M.~Modugno, R.~Saers, C.~Fort, and M.~Inguscio,
Phys.~Rev.~Lett. {\bf 93}, 140406 (2004).

\bibitem{Dalfovo}
F.~Dalfovo, S.~Giorgini, L.~P.~Pitaevskii, and S.~Stringari, 
Rev.~Mod.~Phys. {\bf 71}, 463 (1999).

\bibitem{Huhtamaki} 
J.~A.~M.~Huhtam\"aki, M.~M\"ott\"onen, T.~Isoshima, V.~Pietil\"a, and S.~M.~M.~Virtanen, 
Phys.~Rev.~Lett. {\bf 97}, 110406 (2006).

\bibitem{Munoz}
A.~M.~Mateo and V.~Delgado, Phys.~Rev.~Lett. {\bf 97}, 180409 (2006).

\bibitem{Bogoliubov}
N.~N.~Bogoliubov, J.~Phys.~(Moscow) {\bf 11}, 23 (1947). 

\bibitem{deGennes}
P.~G.~de Gennes, {\it Superconductivity of Metals and Alloys}
(Benjamin, New York, 1966).

\bibitem{Fetter}
A.~L.~Fetter, Ann.~of~Phys.~ {\bf 70}, 67 (1972).

\bibitem{Skryabin}
D.~V.~Skryabin, Phys.~Rev.~A {\bf 63}, 013602 (2000).

\bibitem{Ishino}
S.~Ishino, M.~Tsubota, and H.~Takeuchi, Phys.~Rev.~A {\bf 88}, 063617 (2013).

\bibitem{Wen}
L.~Wen, Y.~Qiao, Y.~Xu, and L.~ Mao, Phys.~Rev.~A. {\bf 87}, 033604 (2013).

\bibitem{Kawaguchi}
Y.~Kawaguchi and T.~Ohmi, Phys.~Rev.~A. {\bf 70}, 043610 (2004).

\bibitem{Mine}
M.~Mine, M.~Okumura, T.~Sunaga, and Y.~Yamanaka, Ann.~ Phys.~(N.Y.) {\bf 322}, 2327 (2007).  

\bibitem{Taylor}
E.~Taylor and E.~Zaremba, Phys.~Rev.~A {\bf 68}, 053611 (2003).

\bibitem{Lundh}
E.~Lundh and H.~M.~Nilsen, Phys.~Rev.~A. {\bf 74}, 063620 (2006).

\bibitem{Nakamura1}
Y.~Nakamura, M.~Mine, M.~Okumura, and Y.~Yamanaka, 
Phys.~Rev.~A {\bf 77}, 043601 (2008).

\bibitem{Takahashi2014}
J.~Takahashi, Y.~Nakamura, and Y.~Yamanaka,
Ann.~Phys. {\bf 347}, 250 (2014).

\bibitem{Pu}
H.~Pu, C.~K.~Law, J.~H.~Eberly, and N.~P.~Bigelow, Phys.~Rev.~A {\bf 59}, 1533 (1999).
\end{thebibliography}
\end{document}